\documentclass[aps,reprint,superscriptaddress,longbibliography,floatfix]{revtex4-2}
\usepackage{amsmath}
\usepackage{graphicx,booktabs,etoolbox}
\usepackage{array}
\usepackage{microtype}
\usepackage{xcolor}
\usepackage[colorlinks=true, linkcolor=blue!50!black, citecolor=blue!50!black, urlcolor=blue!50!black]{hyperref}
\hypersetup{pdfauthor={Akinori Machino et al.}}

\newcommand{\xFundingStatement}{%
This work was supported in part by JST COI-NEXT (Grant No.~JPMJPF2014), JST Moonshot R\&D (Grant Nos.~JPMJMS256L, JPMJMS256F, and JPMJMS256E), MEXT Q-LEAP (Grant Nos.~JPMXS0118068682 and JPMXS0120319794), JST SPRING (Grant No.~JPMJSP2138), and JST ASPIRE (Grant No.~JPMJAP2513).%
}
\newcommand{\xPriorPresentationText}{%
An earlier version of this work was presented as a poster at the 2026 IEEE International Conference on Quantum Computing and Engineering (QCE)%
}

\newcommand{\xPeqValue}{0.415}
\newcommand{\xAlphaStar}{0.2075}

\newcommand{\xNonlocalPc}{0.254}
\newcommand{\xNonlocalPeq}{0.746}
\newcommand{\xSquareModelPeq}{0.5}

\begin{document}
\newcommand{\xTitleLead}{Benchmarking superconducting quantum processors}
\newcommand{\xTitleTail}{with percolating entanglement quality}
\title{\xTitleLead\texorpdfstring{\\}{ }\xTitleTail}
\hypersetup{pdftitle={\xTitleLead\space\xTitleTail}}

\csdef{xAffiliation1}{Graduate School of Engineering Science, The University of Osaka, Toyonaka, Osaka, Japan}
\csdef{xAffiliation2}{Center for Quantum Information and Quantum Biology, The University of Osaka, Toyonaka, Osaka, Japan}
\csdef{xAffiliation3}{QuEL, Inc., Hachioji, Tokyo, Japan}
\csdef{xAffiliation4}{e-trees.Japan, Inc., Hachioji, Tokyo, Japan}
\csdef{xAffiliation5}{Graduate School of Information Science and Technology, The University of Osaka, Suita, Osaka, Japan}
\csdef{xAffiliation6}{RIKEN Center for Quantum Computing, Wako, Saitama, Japan}
\csdef{xAffiliation7}{Graduate School of Engineering, The University of Tokyo, Bunkyo-ku, Tokyo, Japan}
\newcommand{\xFirstAuthorEmail}{akinori.machino@qiqb.osaka-u.ac.jp}
\newcommand{\xLastAuthorEmail}{negoro.makoto.qiqb@osaka-u.ac.jp}
\newcommand{\xAuthorRecords}{%
\xFirstAuthorRecord{Akinori~Machino}{1,2}
\xAuthorRecord{Kazuhisa~Ogawa}{2,5}
\xAuthorRecord{Takefumi~Miyoshi}{2,3,4}
\xAuthorRecord{Hidehisa~Shiomi}{2}
\xAuthorRecord{Shinichi~Morisaka}{2,3}
\xAuthorRecord{Ryo~Matsuda}{1,2}
\xAuthorRecord{Nilton~F.~G.~Filho}{1,2}
\xAuthorRecord{Koichiro~Ban}{5}
\xAuthorRecord{Takafumi~Miyanaga}{2}
\xAuthorRecord{Keisuke~Koike}{4}
\xAuthorRecord{Ryutaro~Ohira}{3}
\xAuthorRecord{Toshi~Sumida}{3}
\xAuthorRecord{Yoshinori~Kurimoto}{3}
\xAuthorRecord{Yuuya~Sugita}{3}
\xAuthorRecord{Yosuke~Ito}{3}
\xAuthorRecord{Yasunari~Suzuki}{6,2}
\xAuthorRecord{Peter~A.~Spring}{6}
\xAuthorRecord{Shiyu~Wang}{6}
\xAuthorRecord{Hiroto~Mukai}{7,6}
\xAuthorRecord{Arvind~Mamgain}{6}
\xAuthorRecord{Shuhei~Tamate}{6}
\xAuthorRecord{Yutaka~Tabuchi}{6,2}
\xAuthorRecord{Yasunobu~Nakamura}{6,7}
\xLastAuthorRecord{Makoto~Negoro}{1,2,5}
}

\newcommand{\xUseAffiliation}[1]{\affiliation{\csuse{xAffiliation#1}}}
\newcommand{\xAuthorRecord}[2]{\author{#1}\forcsvlist{\xUseAffiliation}{#2}}
\newcommand{\xFirstAuthorRecord}[2]{\author{#1}\email[Contact author: ]{\xFirstAuthorEmail}\forcsvlist{\xUseAffiliation}{#2}}
\newcommand{\xLastAuthorRecord}[2]{\author{#1}\email[Contact author: ]{\xLastAuthorEmail}\forcsvlist{\xUseAffiliation}{#2}}
\xAuthorRecords

\makeatletter
\def\@affil@script#1#2#3#4{%
  \@ifnum{#1=\z@}{}{%
    \ifnum#1=5\relax\hspace{1.5em}\else\par\fi
    \begingroup
      \frontmatter@affiliationfont
      \@ifnum{\c@affil<\affil@cutoff}{}{%
        \def\@thefnmark{#1}\@makefnmark
      }%
      \ignorespaces#3%
      \@if@empty{#4}{}{\frontmatter@footnote{#4}}%
      \ifnum#1=4\relax\else\par\fi
    \endgroup
  }%
}
\makeatother

\begin{abstract}
    The capability to create large entanglement resources is a key performance attribute of quantum processors, and the size of the resource is often used as a benchmark of processor scale.
    Recent graph-state experiments have reported entanglement clusters spanning entire chips with hundreds of qubits.
    However, cluster size alone does not capture the entanglement quality of individual couplings or differences in coupling topology.
    Here, we introduce percolating entanglement quality (PEQ), a topology-normalized benchmark based on conditional two-qubit negativity and a topology-specific bond-percolation threshold.
    As an experimental demonstration, we prepare a native graph state on a 64-qubit square-lattice superconducting processor with defects, identify a 52-qubit entanglement cluster, and evaluate the PEQ of the processor.
    Applying PEQ to previously reported chip-wide graph-state measurements on heavy-hex-lattice processors, we show that PEQ differs substantially among processors whose entanglement clusters span the entire chip.
    PEQ provides a single chip-wide score without selecting a best-performing qubit subset or path.
    This score allows processors with different coupling topologies to be compared on a common scale and can inform the design of coupling topologies for scalable chips.
\end{abstract}

\maketitle

\section{Introduction}
\label{sec:introduction}

The capability to create large entanglement resources is a key performance attribute of quantum processors~\cite{cao2023generation,jiang2026one,javadiabhari2025big}. Graph states are one class of such resources, prepared by entangling qubits according to a graph~\cite{hein2004multiparty}. They serve as resources for measurement-based quantum computation (MBQC)~\cite{raussendorf2001oneway} and have also been used to benchmark quantum processors~\cite{hamilton2022entanglement,zander2025benchmarking}. Preparing chip-wide graph states requires coordinated single- and two-qubit operations. Therefore, the resulting entanglement reflects the combined performance of the processor and its control system during simultaneous operation.

Recent studies of chip-wide entanglement in superconducting processors prepare native graph states, whose edges are the physical couplings of the chip, and measure the bipartite entanglement between coupled qubits~\cite{mooney2021whole,kam2024characterization}. For each coupling, they reconstruct the two-qubit state conditioned on measurement outcomes of the neighboring qubits and evaluate its negativity, yielding a spatially resolved entanglement map. These maps reveal chip-wide coupling quality and locate the largest entanglement cluster on the chip.

The size of the largest cluster is often quoted as a measure of processor scale. Cluster size alone, however, does not capture the entanglement quality of individual couplings, because clusters of the same size can consist of couplings with high or low negativity. It also does not capture differences in coupling topology, which determine what fraction of the couplings must remain for a cluster to stay connected. A benchmark for comparing processors with different coupling topologies therefore needs to account for both.

We introduce percolating entanglement quality (PEQ), a topology-normalized benchmark based on graph-state measurements. PEQ measures entanglement quality at the percolation threshold associated with the coupling topology. The evaluation includes all designed couplings on the chip and yields a single score without selecting a best-performing qubit subset or path. This score allows processors with different coupling topologies to be compared on a common scale and can inform the design of coupling topologies for scalable chips.

In this work, we apply PEQ to a 64-qubit square-lattice processor with defects, where we identify a 52-qubit entanglement cluster and obtain a PEQ of 0.42. We compare this result with previously reported chip-wide graph-state measurements on heavy-hex-lattice processors~\cite{mooney2021whole,kam2024characterization}. Although the entanglement clusters of three of these processors span the entire chip, their PEQ values range from 0.20 to 0.74. Additionally, we show by numerical calculations that adding long-range couplings to a square lattice can increase PEQ.

The paper is organized as follows. Section~\ref{sec:framework} presents graph-state characterization and defines PEQ. Section~\ref{sec:results} describes the experiment and reports the measured entanglement map and PEQ. Section~\ref{sec:discussion} compares PEQ across processors and with existing benchmarks, and discusses extensions to other coupling topologies. Section~\ref{sec:conclusion} summarizes the results.

\section{Framework}
\label{sec:framework}

The PEQ evaluation protocol begins with a native graph state prepared across the chip (Fig.~\ref{fig:workflow}). Conditional two-qubit tomography then characterizes entanglement between coupled qubits (Fig.~\ref{fig:measurement}). PEQ summarizes these measurement outcomes in a single chip-wide score that accounts for the coupling topology.

\subsection{Graph-state preparation}

\suppressfloats[t]
\begin{figure}[t]
\centering
\includegraphics[width=\columnwidth]{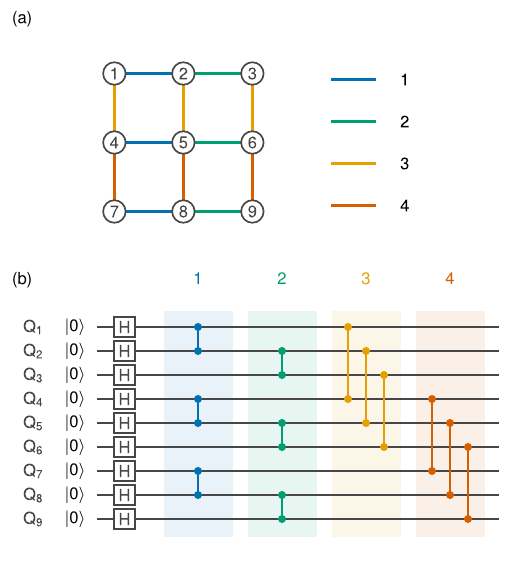}
\caption{Chip-wide graph-state preparation. (a)~Edge coloring of an illustrative $3\times3$ square lattice. Edges of the same color share no vertex, allowing the corresponding CZ gates to operate in parallel. (b)~Corresponding quantum circuit: Hadamard gates act on all qubits initialized in $\lvert0\rangle$, followed by four CZ layers. Connected dots denote CZ gates, and shading groups the gates applied in parallel. The four-layer construction also extends to larger square lattices.}
\label{fig:workflow}
\end{figure}

Let $G=(V,E)$ denote the full coupling graph specified by the chip design, with vertex set $V$ and undirected edge set $E$. Vertices represent qubits, and edges represent designed couplings. The ideal native graph state on $G$ is
\begin{equation}
\lvert G\rangle
=\prod_{(i,j)\in E}\mathrm{CZ}_{ij}
\bigotimes_{k\in V}\lvert+\rangle_k,
\label{eq:graphstate}
\end{equation}
where $\lvert+\rangle=(\lvert0\rangle+\lvert1\rangle)/\sqrt{2}$~\cite{hein2004multiparty}. Hadamard gates prepare all qubits in $\lvert+\rangle$, and controlled-$Z$ (CZ) gates act on every edge in $E$. Assuming that CZ gates sharing a qubit cannot be executed simultaneously, we arrange them in edge-coloring layers. For a square lattice, four CZ layers suffice regardless of the qubit count, as illustrated in Fig.~\ref{fig:workflow}.

\subsection{Negativity measurement}
\label{sec:conditional-negativity}

For a target pair $(i,j)$ of coupled qubits in the chip-wide graph state, we refer to the qubits neighboring the pair as spectators. Following~\cite{mooney2021whole,kam2024characterization}, quantum state tomography is performed on the pair while the spectators are measured in the $Z$ basis. These $Z$ measurements leave each spectator in $\lvert0\rangle$ or $\lvert1\rangle$, disentangled from the remaining qubits. For a graph state, measuring all spectators isolates the target pair from the other qubits. The pair is left in $\mathrm{CZ}_{ij}\lvert+\rangle_i\lvert+\rangle_j$ up to local $Z$ operations, which depend on the spectator outcomes. This two-qubit graph state is maximally entangled and locally equivalent to a Bell state~\cite{hein2004multiparty}.

Because the target-pair state depends on the spectator outcomes, averaging over shots without distinguishing these outcomes hides the entanglement of the pair~\cite{mooney2021whole}. The jointly measured data are therefore grouped by spectator outcome pattern $\boldsymbol{s}$ in post-processing. For each pattern, a two-qubit density matrix $\rho_{ij\mid \boldsymbol{s}}$ is reconstructed from the corresponding target-pair data.

The two-qubit entanglement is quantified by the negativity~\cite{vidal2002computable},
\begin{equation}
\mathcal{N}(\rho_{ij\mid \boldsymbol{s}})=
\frac{\left\lVert\rho_{ij\mid \boldsymbol{s}}^{\top_i}\right\rVert_1-1}{2},
\label{eq:negativity}
\end{equation}
where $\top_i$ denotes partial transposition on qubit $i$ and $\lVert\cdot\rVert_1$ is the trace norm. Two-qubit negativity lies between zero and $1/2$, with the upper bound attained by a maximally entangled state. A positive value indicates entanglement.

In a real device, noise can lead to different negativities for the conditional states. Following~\cite{mooney2021whole,kam2024characterization}, we assign the largest conditional negativity of each pair to the corresponding edge of the coupling graph. The edge score is thus
\begin{equation}
\mathcal{N}_{ij}=\max_{\boldsymbol{s}}\mathcal{N}(\rho_{ij\mid \boldsymbol{s}}).
\label{eq:bestbranch}
\end{equation}

\begin{figure}[tbp]
\centering
\includegraphics[width=\columnwidth]{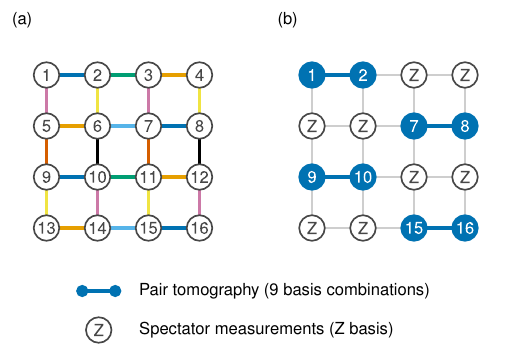}
\caption{Parallel negativity measurement. (a)~Periodic strong edge coloring of a $4\times4$ square lattice, partitioning its edges into eight measurement groups. (b)~Four target pairs in the blue group measured in parallel. The state of each pair is reconstructed from quantum state tomography conditioned on the $Z$-measurement outcomes of its spectators.}
\label{fig:measurement}
\end{figure}

Target pairs are partitioned into measurement groups by a strong edge coloring, in which two edges receive different colors if they share a vertex or are joined by another edge~\cite{fouquet1983strong}. Pairs in the same group therefore share no qubits and have no direct couplings between them (Fig.~\ref{fig:measurement}). For an ideal graph state, measuring the spectators of all pairs in a group leaves a tensor product of maximally entangled two-qubit states, one for each pair. The pairs in each group can therefore be measured in parallel even if they share a $Z$-measured spectator~\cite{kam2024characterization}. With a bounded number of neighbors per qubit, a fixed number of measurement groups suffices regardless of chip size. The periodic square-lattice coloring illustrated in Fig.~\ref{fig:measurement} uses eight groups. Nine combinations of Pauli measurement bases per group give 72 distinct measurement circuits, independent of lattice size.

\subsection{Entanglement yield and PEQ}

Previous studies report the number of qubits in the largest entanglement cluster~\cite{mooney2021whole,kam2024characterization}. To compare chips of different sizes, we normalize this number by the total number of qubits on the chip and define the entanglement yield (EY) as
\begin{equation}
\mathrm{EY}(\alpha)=\frac{\lvert V_{\max}(\alpha)\rvert}{\lvert V\rvert},
\label{eq:ey}
\end{equation}
where $\lvert\cdot\rvert$ denotes the number of elements in a set and $V_{\max}(\alpha)$ is the vertex set of a largest connected cluster after retaining only edges with negativity above a cutoff $\alpha$. All qubits on the chip, including isolated ones, are counted in $\lvert V\rvert$. At zero cutoff, where every coupling with positive negativity is retained, EY gives the fraction of qubits in the largest entanglement cluster identified in the prepared chip-wide graph state. It therefore serves as a yield metric for a quantum processor.

EY measures entanglement cluster size but does not distinguish couplings with high negativity from those with low negativity. Here, we apply percolation theory to quantify the entanglement quality that can support connectivity across the lattice~\cite{saberi2015recent}. When each edge of an infinite lattice is retained independently with probability $p$, a cluster spanning the lattice exists only if $p$ exceeds the bond-percolation threshold $p_{\mathrm c}$, whose value depends on the coupling topology~(Appendix~\ref{app:percolation-thresholds}). We use $p_{\mathrm c}$ as the reference fraction of edges for determining the negativity cutoff. Writing $\mathcal{N}_e$ for the negativity score on an edge $e$, we define the retained edge set $E(\alpha)$ and its occupation fraction $p(\alpha)$ as
\begin{gather}
E(\alpha)=\{e\in E:\ \mathcal{N}_e>\alpha\},
\label{eq:retained}\\
p(\alpha)=\frac{\lvert E(\alpha)\rvert}{\lvert E\rvert},
\label{eq:occupation}
\end{gather}
where all edges, i.e., designed qubit--qubit couplings, are counted in the denominator.

To obtain PEQ, we count couplings in descending order of negativity until their fraction reaches or exceeds $p_{\mathrm c}$. The negativity of the last coupling counted is denoted as $\alpha^*$. PEQ is defined as
\begin{equation}
\mathrm{PEQ}=2\alpha^*.
\label{eq:peq}
\end{equation}
The factor of two normalizes PEQ to a zero-to-one scale. PEQ reaches unity when all couplings in the required fraction have the maximal negativity of $1/2$. If the required fraction cannot be reached by using couplings with nonzero negativity, PEQ is zero.

PEQ characterizes coupling quality using a topology-specific percolation threshold as a reference, while EY quantifies the size of the connected entanglement cluster. Although the threshold differs among two-dimensional lattices, the critical behavior at the threshold is the same~\cite{saberi2015recent}, which allows different coupling topologies to be compared at their respective thresholds. In MBQC, a cluster state with randomly missing qubits or bonds remains a universal resource above the percolation threshold~\cite{browne2008phase,kieling2007percolation}. PEQ does not test whether the retained edges are connected on the finite chip, and the percolation threshold does not guarantee a spanning cluster in a finite device. These quantities are intended to benchmark processor-wide entanglement rather than certify genuine multipartite entanglement of the graph state~\cite{cao2023generation,zander2025benchmarking}.

\begin{figure*}[t]
\centering
\includegraphics[width=\textwidth]{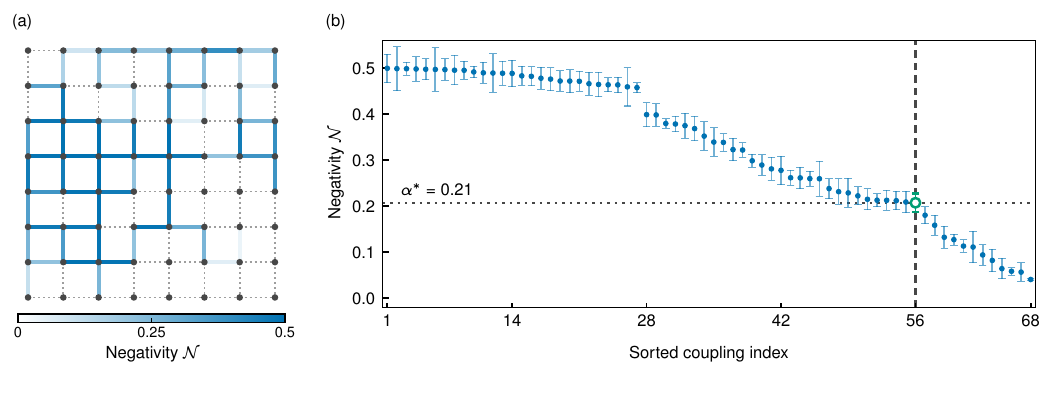}
\caption{Entanglement map of the 64-qubit square-lattice processor~\cite{machino2026peq}. (a)~Measured negativities $\mathcal{N}$, with gray dotted links for defect couplings without negativity data. (b)~Negativities in descending order, with one-standard-deviation error bars. PEQ uses all 112 designed couplings. At $p_{\mathrm c}=0.5$, the 56th coupling determines the cutoff $\alpha^*$, marked with a green circle, and PEQ $=0.42$.}
\label{fig:definition}
\end{figure*}
\section{Experimental Results}
\label{sec:results}

\subsection{Preparation and acquisition}

The experiment uses a 64-qubit fixed-frequency superconducting quantum processor, and microwave control is provided by the QuBE/Qubex system~\cite{machino2026qubex}. The coupling graph of the processor is an $8\times8$ square lattice with 112 edges. The corresponding infinite square lattice has a bond-percolation threshold of $p_{\mathrm c}=0.5$~\cite{kesten1980critical}. Qubits and couplings for which the required operations could not be calibrated are treated as defects and excluded from graph-state preparation. The occupation fraction $p(\alpha)$ is calculated with all 112 edges in the denominator. Device parameters and gate fidelities are summarized in Appendix~\ref{app:device-calibration}.

Preparation initializes the participating qubits in $\lvert+\rangle$ and applies CZ gates in layers of disjoint pairs. Each CZ gate is implemented using a cross-resonance-based CNOT gate, with Hadamard gates on the target qubit before and after the CNOT. We perform local conditional quantum state tomography as described in Sec.~\ref{sec:conditional-negativity}. The eight measurement groups described there cover the target pairs, each using the nine target-pair basis combinations in $\{X,Y,Z\}^{\otimes2}$ with 3000 shots per combination. Neighboring spectators are measured in $Z$, and their outcomes are used to group the target-pair data.

\subsection{Reconstruction and uncertainty}

We apply readout-error mitigation to the joint measurement counts before conditioning on spectator outcomes. We reconstruct each density matrix by a least-squares fit to the measured Pauli expectation values, subject to positive-semidefinite and unit-trace constraints. Negativity is calculated from each reconstructed matrix, and the edge score is assigned according to Eq.~\eqref{eq:bestbranch}.

We estimate the uncertainty in negativity by resampling the tomography data. We report the sample standard deviation of the negativities obtained from 200 Dirichlet resamples using linear density-matrix reconstruction. For PEQ, we propagate the negativity uncertainty of the coupling that determines $\alpha^*$ through Eq.~\eqref{eq:peq}.

\subsection{Measured entanglement map and PEQ}

Figure~\ref{fig:definition} shows the spatial map of the measured negativities and their distribution in descending order. At zero negativity cutoff, the largest connected component of the entanglement graph contains 52 qubits, giving $\mathrm{EY}(0)=52/64=0.8125$. All 68 measured couplings have negativity estimates exceeding twice their resampling standard deviations, giving an occupation fraction $p(0)=68/112=0.607$, above the square-lattice percolation threshold of 0.5.

At the square-lattice percolation threshold, 56 of the 112 couplings must be retained. The ordered data in Fig.~\ref{fig:definition}(b) give $\alpha^*=\xAlphaStar$ and PEQ $=\xPeqValue$. Propagating the one-standard-deviation error of the negativity of the 56th coupling gives the rounded values $\alpha^*=0.21\pm0.02$ and PEQ $=0.42\pm0.04$.

Here, $\mathrm{EY}(0)=0.81$ indicates that 81\% of the qubits on the chip belong to a single connected entanglement cluster. PEQ $=0.42$ characterizes the entanglement quality that can support percolation on the square lattice. These complementary measures describe the extent and quality of chip-wide entanglement and provide a basis for comparison across processors.

\section{Discussion}
\label{sec:discussion}
\subsection{Comparison across coupling topologies}

\begin{figure*}[t]
\centering
\includegraphics[width=\textwidth]{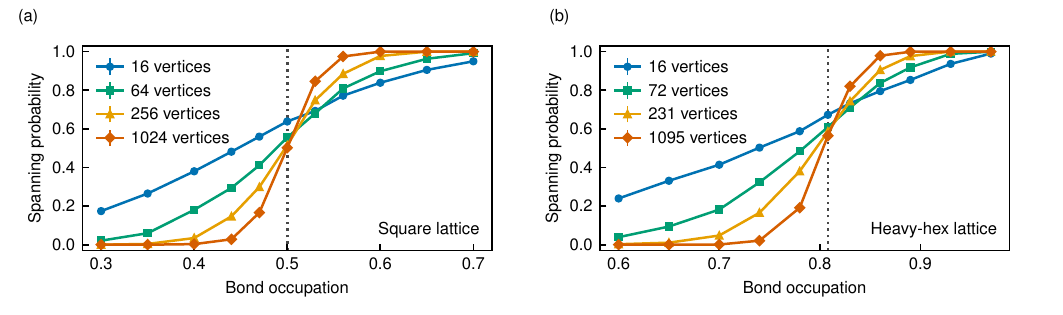}
\caption{Finite-size spanning probabilities on (a)~square and (b)~heavy-hex lattices. Legends give the number of vertices. Error bars show the standard error and are smaller than the markers. Vertical dotted lines mark the infinite-lattice thresholds.}
\label{fig:finite-size}
\end{figure*}

By defining a common measure based on topology-specific percolation thresholds, PEQ enables comparisons of chip-wide coupling quality across different coupling topologies. Table~\ref{tab:peq-comparison} compares our square-lattice processor with four heavy-hex-lattice processors for which previous chip-wide graph-state experiments reported the negativity of each coupling~\cite{mooney2021whole,kam2024characterization}. The values reflect each processor at the time of its experiment. The square and heavy-hex lattices have bond-percolation thresholds $p_{\mathrm c}$ of 0.5 and approximately 0.808, respectively~(Appendix~\ref{app:percolation-thresholds} and Fig.~\ref{fig:finite-size}).

\suppressfloats[t]
\begin{table}[t]
\centering
\caption{Entanglement yield~(EY) at zero negativity cutoff and percolating entanglement quality~(PEQ) across processors. Counts cover the full chip design.}
\label{tab:peq-comparison}
\renewcommand{\thempfootnote}{\fnsymbol{mpfootnote}}
\newcommand{\xComparisonNoteMark}{%
\footnotemark[2]}
\setlength{\tabcolsep}{1.8pt}
\begin{tabular*}{\columnwidth}{@{\extracolsep{\fill}}llrrrr@{\hspace{0.4em}}}
\toprule
Processor & Topology & Qubits & Bonds & EY & PEQ\\
\midrule
This work & square & 64 & 112 & 0.81 & 0.42\\
\texttt{ibmq\_rochester}~\cite{mooney2021whole} & heavy-hex & 53 & 58 & 1.00 & 0.20\rlap{\xComparisonNoteMark}\\
\texttt{ibmq\_manhattan}~\cite{mooney2021whole} & heavy-hex & 65 & 72 & 1.00 & 0.74\rlap{\xComparisonNoteMark}\\
\texttt{ibm\_washington}~\cite{kam2024characterization} & heavy-hex & 127 & 144 & 1.00 & 0.66\rlap{\xComparisonNoteMark}\\
\texttt{ibm\_seattle}~\cite{kam2024characterization} & heavy-hex & 433 & 504 & 0.96 & 0.39\rlap{\xComparisonNoteMark}\\
\bottomrule
\end{tabular*}
\footnotetext[2]{Calculated from the negativity values extracted from the readout-error-mitigated plots in~\cite{mooney2021whole,kam2024characterization}.}
\end{table}

PEQ provides information about coupling quality beyond the cluster coverage measured by EY. Our square-lattice processor has a lower EY than \texttt{ibmq\_rochester}~(0.81 versus 1.00), but a higher PEQ (0.42 versus 0.20). Even with the same heavy-hex topology and EY $=1.00$, \texttt{ibmq\_rochester} and \texttt{ibmq\_manhattan} have substantially different PEQ values, at 0.20 and 0.74, respectively.

The 433-qubit \texttt{ibm\_seattle} processor is significantly larger than the 127-qubit \texttt{ibm\_washington} processor but has a lower PEQ~(0.39 versus 0.66). PEQ reflects the quality of entanglement generated during chip-wide operation. The quality is degraded by limited qubit coherence, imperfect gates, and crosstalk during simultaneous operations~\cite{kam2024characterization}.

\subsection{Relation to existing quantum benchmarks}

Quantum-processor benchmarks characterize different aspects of performance and can evaluate different parts of a device. We compare what each benchmark measures, the information it provides, and which part of the chip it evaluates. Table~\ref{tab:benchmarks} summarizes what each benchmark measures.

\suppressfloats[t]
\begin{table}[t]
\centering
\caption{Quantities characterized by quantum-processor benchmarks.}
\label{tab:benchmarks}
\begin{tabular*}{\columnwidth}{@{\extracolsep{\fill}}lp{0.78\columnwidth}@{}}
\toprule
Metric & What it measures\\
\midrule
\addlinespace[4pt]
QV & \raggedright Executable circuit width and depth\tabularnewline
\addlinespace[2pt]
EPLG & \raggedright Gate error during parallel operation\tabularnewline
\addlinespace[2pt]
QUOPS & \raggedright Executable circuit size and effective speed\tabularnewline
\addlinespace[2pt]
XEB & \raggedright Fidelity of random circuit execution\tabularnewline
\addlinespace[2pt]
PEQ & \raggedright Entanglement quality at the percolation threshold\tabularnewline
\addlinespace[0pt]
\bottomrule
\end{tabular*}
\end{table}

\begin{description}
\item[QV]
Quantum volume (QV) is $2^m$, where $m$ is the largest width and matching depth of random model circuits that satisfy a heavy-output-generation criterion with statistical confidence~\cite{cross2019validating}. The score combines the effects of gate operations, compilation, routing, and measurement, providing a test of circuit execution. On a large chip, the tested circuits may use only a small subset of qubits~\cite{mckay2023benchmarking}.

\item[EPLG]
Error per layered gate (EPLG) summarizes errors during parallel execution by normalizing the layer fidelity obtained from simultaneous direct randomized benchmarking~\cite{mckay2023benchmarking,lozano2026parameter}. The measurements estimate the error of each gate during parallel operation, including crosstalk effects. The protocol applies to connected groups of qubits; the studies in~\cite{mckay2023benchmarking,lozano2026parameter} evaluate selected one-dimensional qubit chains. A chain-based EPLG value characterizes the qubits and gates in the chain rather than all designed couplings on the chip.

\item[QUOPS]
The quantum universal operation performance system (QUOPS) varies circuit width and depth independently within prescribed shape constraints, whereas QV uses square model circuits~\cite{proctor2026benchmarking}. QUOPS reports the largest executable circuit size as its score and the effective rate of quantum operations in the circuit shape. On a large chip, the reported score and rate may characterize only a selected subset of qubits~\cite{proctor2026benchmarking}.

\item[XEB]
Cross-entropy benchmarking (XEB) estimates the fidelity of random quantum circuits by comparing measured bitstrings with classically computed output probabilities~\cite{boixo2018characterizing,arute2019quantum}. Applied to a whole chip, it gives a circuit fidelity that reflects the accumulated errors of all gates and measurements. Its verification requires classical simulation of the circuits, which becomes impractical at large qubit counts and depths.

\item[PEQ]
Percolating entanglement quality (PEQ) measures entanglement quality at the percolation threshold associated with the coupling topology. The evaluation always covers all designed couplings on the chip, including defects. The graph-state preparation probes the combined performance of the calibrated processor and its control system during chip-wide parallel operation. The measurement uses a fixed number of circuits, 72 for a square lattice, regardless of the number of qubits (Sec.~\ref{sec:conditional-negativity}). The score characterizes the prepared graph state and does not directly measure gate fidelities or the depth of circuits that can be executed reliably.
\end{description}

QV, EPLG, and QUOPS can characterize the performance of selected qubit subsets, and chip-wide XEB requires classical simulation of the circuits. PEQ evaluates all designed couplings of the fabricated chip, including defects, with a fixed number of circuits. PEQ is therefore a useful reference for comparing fabricated processors and informing improvements in chip fabrication.

The same percolation-based rule can be applied to quantities other than negativity. For example, ranking the two-qubit gate fidelities obtained from layer-fidelity measurements~\cite{mckay2023benchmarking,lozano2026parameter} and taking the value at the percolation threshold defines a percolating gate fidelity. This quantity characterizes gate quality across the full chip and complements EPLG, whereas PEQ characterizes the entanglement of the prepared state.

\subsection{Extension to other coupling topologies}

PEQ is not limited to the square and heavy-hex lattices and can be defined for any coupling topology. For a periodic coupling pattern, the bond-percolation threshold of the corresponding infinite lattice determines the required fraction of retained couplings. When the threshold is not known analytically, it can be estimated by Monte Carlo simulations on progressively larger patches.

\begin{figure*}[!tp]
\centering
\includegraphics[width=\textwidth]{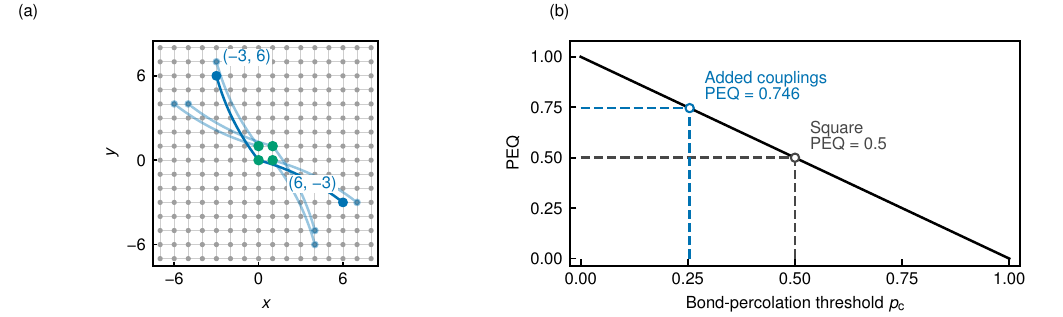}
\caption{Periodic long-range couplings and PEQ comparison. (a)~Blue curves show added couplings for the four green sites of a $2\times2$ cell, which is periodically repeated to compose a lattice for the gross code. (b)~Large-lattice-limit PEQ of the square lattice and the lattice with added long-range couplings, assuming a uniform negativity distribution on $[0,1/2]$.}
\label{fig:nonlocal-model}
\end{figure*}

As an example with a lower bond-percolation threshold than the square lattice, we consider a periodic coupling topology corresponding to the gross quantum error-correcting code~\cite{bravyi2024high}. The topology is obtained by adding long-range couplings to a square lattice [Fig.~\ref{fig:nonlocal-model}(a)]. Multilayer superconducting wiring has been proposed for the coupling topology~\cite{mathews2026placing}. For the lattice with additional couplings, we estimate $p_{\mathrm c}\approx\xNonlocalPc$, compared with 0.5 for the square lattice (Appendix~\ref{app:percolation-thresholds}).

To compare PEQ between the two coupling topologies, we independently assign each coupling a negativity drawn from a uniform distribution on $[0,1/2]$. In the large-lattice limit, the fraction of couplings with negativity above $\alpha$ is $1-2\alpha$, which equals $p_{\mathrm c}$ at $\alpha^*=(1-p_{\mathrm c})/2$. This gives $\mathrm{PEQ}=2\alpha^*=1-p_{\mathrm c}$, yielding \xSquareModelPeq{} for the square lattice and \xNonlocalPeq{} for the lattice with additional long-range couplings [Fig.~\ref{fig:nonlocal-model}(b)]. The lower percolation threshold allows a higher negativity cutoff, resulting in a higher PEQ despite the unchanged negativity distribution.

\begin{figure*}[!tp]
\centering
\includegraphics[width=\textwidth]{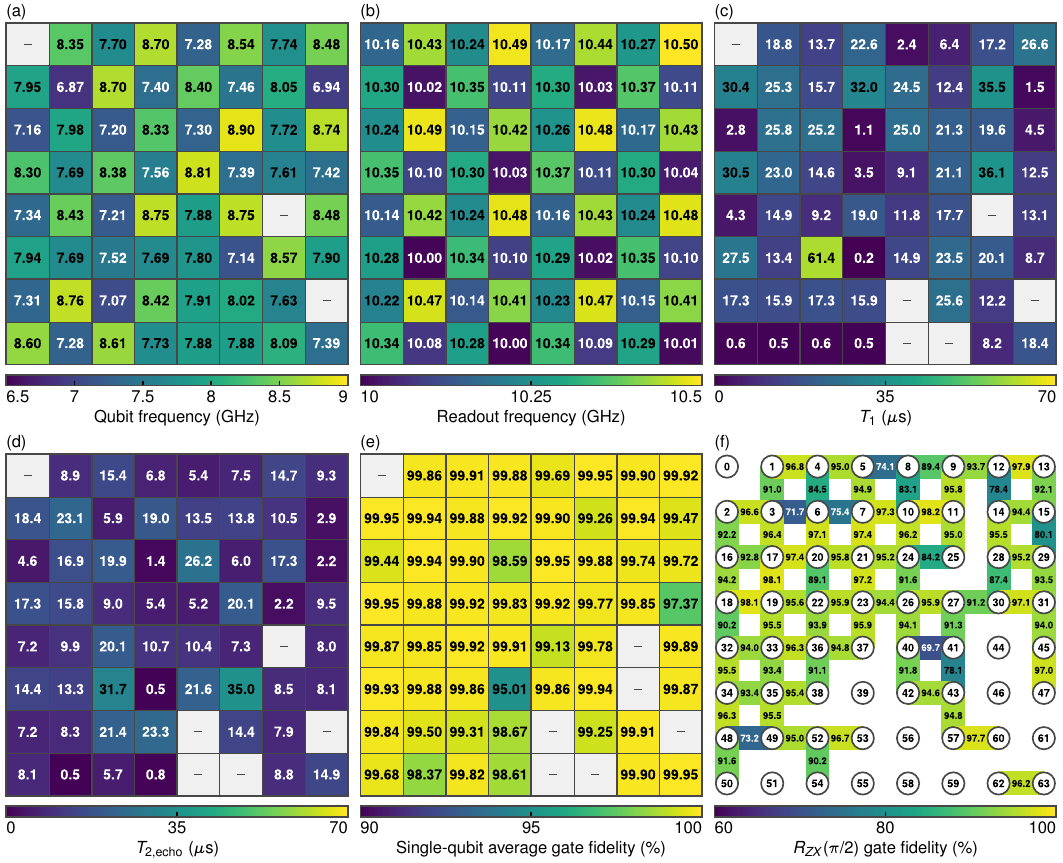}
\caption{Device parameters and gate fidelities on the $8\times8$ lattice: (a)~qubit frequency, (b)~readout frequency, (c)~$T_1$, (d)~$T_{2,\mathrm{echo}}$, (e)~single-qubit average gate fidelity, and (f)~$R_{ZX}(\pi/2)$ gate fidelity. Gray cells indicate unavailable values or failed calibrations. Couplings are shown only where gate-fidelity values are available.}
\label{fig:device-calibration-maps}
\end{figure*}

\section{Conclusion}
\label{sec:conclusion}

We have introduced PEQ, a single zero-to-one score for the quality of entanglement generated across a quantum processor. Using the percolation threshold associated with the coupling topology, PEQ provides a common basis for comparing different chip designs. On a 64-qubit square-lattice processor, we identified a 52-qubit entanglement cluster and obtained a PEQ of 0.42.

Although \texttt{ibmq\_rochester} and \texttt{ibmq\_manhattan} both have entanglement clusters spanning the entire chip, applying PEQ to their graph-state measurements gives substantially different values of 0.20 and 0.74, respectively. Numerical calculations further show that adding long-range couplings to a square lattice raises PEQ from \xSquareModelPeq{} to \xNonlocalPeq{} for a uniform negativity distribution. PEQ thus allows processors with different coupling topologies to be compared on a common scale and can inform the design of coupling topologies for scalable chips.

\begin{acknowledgments}
\xFundingStatement\par
The authors acknowledge the use of GPT-6 Astra (OpenAI) and Claude Fable 5.2 (Anthropic) to assist in preparing the manuscript, reviewing the literature, and developing experimental scripts. The authors reviewed and verified the AI-assisted content and take full responsibility for the final manuscript.
\par\xPriorPresentationText~\cite{machino2026peq}.\end{acknowledgments}
\section*{Data Availability}
The data and analysis code supporting this work are openly available on Zenodo~\cite{machino2026peqdata}.

\appendix
\section{Percolation calculations}
\label{app:percolation-thresholds}

In bond percolation, each lattice edge (bond) is independently retained with probability $p$, while all vertices are kept~\cite{saberi2015recent}. At low $p$, the retained bonds form only finite connected clusters. For $p>p_{\mathrm c}$, an infinite connected cluster exists on the infinite lattice. The threshold $p_{\mathrm c}$ depends on the coupling topology.

Bond percolation on the infinite square lattice has $p_{\mathrm c}=0.5$~\cite{kesten1980critical}. The honeycomb lattice, in which each vertex has three neighbors instead of four, has a higher threshold, $p_{\mathrm c}^{\mathrm{honeycomb}}=1-2\sin(\pi/18)\approx0.653$~\cite{sykes1964exact}. The heavy-hex-lattice coupling graphs~\cite{mooney2021whole,kam2024characterization} can be obtained by replacing each honeycomb-lattice edge with two bonds in series. Connecting the original endpoints requires both bonds to be occupied, which occurs with probability $p^2$. Setting $p^2=p_{\mathrm c}^{\mathrm{honeycomb}}$ gives
\begin{equation}
p_{\mathrm c}^{\mathrm{heavy\text{-}hex}}
=\sqrt{1-2\sin(\pi/18)}\approx0.808.
\end{equation}

Figure~\ref{fig:finite-size} shows the probability that a cluster spans finite lattice patches from left to right, estimated from 2000 random bond configurations at each occupation fraction. As the number of vertices increases, the transitions become sharper and approach the infinite-lattice thresholds. Full simulation settings are provided in the supporting data and code.

The lattice in Fig.~\ref{fig:nonlocal-model}(a) has two additional long-range couplings per vertex on top of a square lattice. Its periodic $2\times2$ coupling pattern corresponds to that of the gross code~\cite{bravyi2024high}. Extrapolating finite-size spanning thresholds using two-dimensional percolation scaling~\cite{newman2001fast,ziff2002convergence}, we estimate $p_{\mathrm c}\approx\xNonlocalPc$. The lattice definition, simulation settings, and extrapolation procedure are provided in the supporting data and code.

\section{Device characterization}
\label{app:device-calibration}

The device characterization in Fig.~\ref{fig:device-calibration-maps} was performed separately from the graph-state experiment during the same cooldown, and its values do not necessarily reflect the condition of the device on the day of that experiment. In particular, the couplings with gate-fidelity values in Fig.~\ref{fig:device-calibration-maps}(f) do not coincide with the couplings used for graph-state preparation in Fig.~\ref{fig:definition}. The gate fidelities in Fig.~\ref{fig:device-calibration-maps} were measured for each gate individually, whereas the graph-state preparation applies the gates in parallel. The median $T_1$, $T_{2,\mathrm{echo}}$, and $T_2^*$ are $15.9$, $9.4$, and $6.9\,\mu\mathrm{s}$, respectively. The median single-qubit average gate fidelity is $99.87\%$, as determined by randomized benchmarking of Clifford gates. Interleaved randomized benchmarking gives a median two-qubit gate fidelity of $94.8\%$ across 77 couplings.

\end{document}